\documentclass{article}

\usepackage[english]{babel}

\usepackage[letterpaper,top=2cm,bottom=2cm,left=3cm,right=3cm,marginparwidth=1.75cm]{geometry}

\usepackage{amsmath}
\usepackage{graphicx}
\usepackage[colorlinks=true, allcolors=blue]{hyperref}
\usepackage{biblatex} %Imports biblatex package
\usepackage{url}

\title{Valuation of Music Catalogs}
\author{Ivan Kosyuk and Sasha Stoikov}

\begin{document}
\maketitle

\begin{abstract}
%The trading of music rights has become more accessible and liquid in the past few years. As those assets generate stable revenues linked to streaming, a question emerges: What is the most effective way to forecast cashflows and establish an appropriate price for the asset? 

We propose a risk neutral approach to forecast the cashflows of music catalogs, based on historical revenue data. We use a discounted cashflows formula to produce reasonable ranges of multipliers for these assets, based on the age of the catalog, the last-twelve-months revenue and the duration of the contract. We compare the multipliers implied by the cashflows of top, median and bottom performing songs on the Royalty Exchange platform. We find that ask prices are close to the multipliers justified by median song cashflows. The best bids are near the multipliers justified by the bottom decile of song cashflows. %We focus on the 10 year contracts traded on Royalty Exchange and look at the cashflows they have generated over time. We show that in the past 10 years, cashflows from older catalogs have grown significantly faster than newer catalogs. This translates into higher multipliers, something that we observe from past transactions on Royalty Exchange.

%While our final result is limited, the end product can serve as a useful reference in deciding on the appropriate pricing for music assets as well as indicating outliers for further research.

\end{abstract}

\section{Introduction}

Nowadays, when new music is released, streaming revenues look small relative to what the sale of CDs used to generate. This has had a profound impact on the timing and magnitude of musicians' revenues \cite{Hesmondhalgh}. The slow trickle of cashflows in the early years after a release may be helpful in predicting the long term streaming revenues for a song or album. This has opened the possibility for financial instruments that allow investors to participate in the future cashflows of music, an activity that has grown to an estimated US\$5bn in music rights transactions in 2021 \cite{goldman}. The structure of these instruments have been debated from a legal perspective \cite{garcia}. From a practical perspective, musicians may want to sell their rights to receive payments upfront, based on the revenue they are expected to generate over a future time period. The money raised this way can provide a lump sum to fund music production, marketing and help artists diversify their assets. 

At the fundamental level, the value of these rights depends on: 
\begin{itemize}
\item The Last Twelve Month (LTM) revenues. Prices are usually quoted in terms of multipliers of the LTM. 
\item The duration $d$ of the rights. This may be a fixed horizon, say 10 or 30 years, or life-of-rights assets that last 70 years after the death of the artist. 
\item The discount rate $r$ used to discount cashflows. There has been much debate about what the discount rate should be \cite{ashford},\cite{massarsky}, \cite{ingham} and \cite{kroll}. In this paper we fix this rate at 10\%. 
\item The expected future cashflows $\hat{C}_{t+i}$ , where $t$ is a value-weighted average of the ages of the songs in a catalog and $1 \leq i \leq d$ are the future dates of the cashflows. Other metadata such as the the genre, the sources of revenue, the nature of the rights (recording or publishing copyrights), the artist's language, country and gender may also be considered. The overall outlook of the the music streaming business may also be an important factor (see the latest "Music in the Air" report by Goldman Sachs \cite{goldman}).
\end{itemize}
The fair price can be expressed as a sum of expected discounted cashflows,
$$P_{t}^{d}=\sum_{i=1}^{d}   \frac{\hat{C}_{t+i}}{(1+r)^i}
$$
where the main challenge is estimating the cashflows $\hat{C}$. Cashflows are notoriously hard to estimate, as hits tend to emerge in very unpredictable ways. Effectively, rights buyers are buying a form of 
high dividend stock, which pays steady cashflows most of the time and occasionally offers significant upside when a positive event occurs. 

In Section 2, we illustrate this cashflow behavior with revenue data from Royalty exchange. We compute 3 revenue growth curves, for the top songs (top decile), median songs and bottom songs (lowest decile), based on cashflows collected on Royalty Exchange. 
In Section 3, we use discounted cashflow formulas using these top, median and bottom revenue curves to derive a range of reasonable multipliers for fixed duration assets. %We find that these multipliers are close to those observed in the market data on Royalty Exchange, for asset of various duration and ages.

\section{Revenue Data}
We collected data on catalogs on Royalty exchange, where each item may comprise of a single song, albums, or collections of songs (referred to as "assets"). For each asset, we had
a history of either monthly or quarterly cashflows as well as a value-weighted age of the asset (referred to as the "dollar age").

We processed the data in two ways - first comparing the age of the oldest cashflow with the dollar age of the assets, and then filtering dollar age outliers. In the first step, we transformed all cashflows into annual cashflows and threw away songs with years of zero revenue. In the second step, we retained songs with dollar ages within a 30\% margin of the age of the oldest cashflow. This ensured that we only work with assets comprising of one song or collections of songs released at approximately the same time.

%That is how we arrive at our data points of interest:
%\begin{itemize}
%\item $D^i$ is a current dollar age of the song $i$
%\item 
%\end{itemize}

%With this data, we calculate a multiplier for an asset of dollar age $d$ and year length $y_c$. 

%We do so by estimating the fraction of last year's cashflows (LTM) each of the next $d$ years will be. 
Since asset prices are typically quoted in multipliers of the $LTM=C_t$, we express the multiplier of asset $i$ with age $t$ and duration $d$ as 
$$M_{t}^d=\frac{P_{t}^{d}}{C_t}=\sum_{i=1}^{d}   \frac{\hat{C}_{t+i}}{C_t (1+r)^i} =\sum_{i=1}^{d}   \frac{\hat{S}_{t,i}}{(1+r)^i}
$$
where $$\hat{S}_{t,i}=\frac{\hat{C}_{t+i}}{C_t}$$ is an estimate of revenues in year $t+i$ expressed as a share of the revenues in year $t$.
%Define $C_k^i$ is a cashflow for song $i$ on year $k$.

To do so, we:
\begin{itemize}
\item Gather all the songs in our data of dollar age at or above $t+i$
%\item Pick cashflow ($C_d^i$) as $LTM^i$ for song $i$ 
\item %For each year $y$ in a range from 1 to $y_c$, we picked 
Compute an observed share that is calculated for each song $j$ as $S_{t,i}^j=\frac{C^j_{t+i}}{C_t^j}$ 

%if the cashflow for that year exists for that song
 %For each year y we pick the desired estimate based on observed shares for that year. 
\item
For each pair of dates $t$ and $i$, %The method for picking the desired estimate can be chosen freely, but we used three: 
compute the median, 90th percentile, and 10th percentile of the variable $S_{t,i}^j$ over all songs $j$. %All methods produce a single number for an estimated share for year y. We found that using the mean instead of the median was less sensitive to outliers. 
The result of these computations is: %$S^{50}_{t,i}$ for median computation, $S^{90}_{t,i}$ for 90th percentile and $S^{10}_{t,i}$ for 10th percentile.
\begin{itemize}
\item$S^{50}_{t,i}$ - revenue share for year $i$ for asset of dollar age $t$ using median approach 
\item$S^{90}_{t,i}$ - revenue share for year $i$ for asset of dollar age $t$ using 90th percentile approach 
\item$S^{10}_{t,i}$ - revenue share for year $i$ for asset of dollar age $t$ using 10th percentile approach 

\end{itemize}

\end{itemize}
%After establishing estimated shares, we sum up the discounted values to get the multiplier. 

%We compare these revenue curves for songs of 1, 3, 5 and 7 year old songs and find that for top performing songs, older catalogs have cashflows that grow faster than newer catalogs. 

\begin{figure}[h!]
\centering
\includegraphics[width=.5\textwidth]{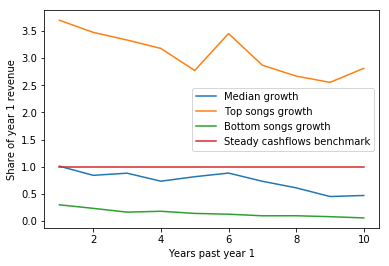}\hfill
\includegraphics[width=.5\textwidth]{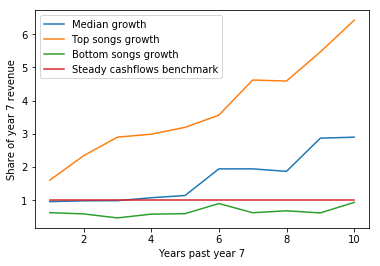}
\caption{\label{fig:cashflows} Top, median and bottom cashflows for songs of age 1 and 7 years 
 }

\end{figure}

These normalized revenue shares are displayed in Figure \ref{fig:cashflows} for $t=1,7$.
We find that the cashflows of very new songs (dollar age = 1) decay for the top, median and bottom performing songs. However, for older songs (dollar age = 7) only bottom performing songs decay, while top performing songs grow significantly. This illustrates the idea that investors who pick top performing songs can see significant upside, when the dollar age is higher.
\pagebreak

\section{Computing multipliers}
From these cashflow growth and decay rates, we may compute the multiplier for top decile assets

$$M^{90}_{t,d}=\sum_{i=1}^{d}   \frac{S^{90}_{t,i}}{(1+r)^i}
$$
median assets
$$M^{50}_{t,d}=\sum_{i=1}^{d}   \frac{S^{50}_{t,i}}{(1+r)^i}
$$
and bottom assets
$$M^{10}_{t,d}=\sum_{i=1}^{d}   \frac{S^{10}_{t,i}}{(1+r)^i}
$$
and compare them to bid and ask prices in the market.
%of dollar age $t$ with a contract of duration $d$ 
%$$M^{50}_{t,d}=\sum_{i=1}^{d}   \frac{S^{50}_{t,i}}{(1+r)^i}
%$$
%using median approach and multipliers for 10th and 90th percentiles are calculated in a similar fashion, but using $S^{90}_{ti}$ and $S^{10}_{ti}$.
%We plot multipliers in Figure \ref{fig:cashflows} for songs of dollar ages 1, 3, 5 and 7. 

%\subsection{Multiplier Market Data}

To evaluate the fit of our model, we observed multipliers from real buyers and sellers on the Royalty Exchange for assets of duration 10 years or less. %To get a broader perspective of the market behavior, f
The market mechanism on Royalty Exchange is similar to that of Ebay, where a single sellers may post an ask price, while multiple buyers compete with their bids. For each song, we gathered the implied multiplier of the most recent bid (by dividing the most recent bid by revenue of the last year) as well as the ask multiplier the seller offers as a "Buy Now" option. As expected, bids are lower than asks, but we added an additional filter - removed observations where the bid multipliers are less than half the ask multipliers. %For such instances, we assumed that either or both values are far away from reasonable and are not an adequate evaluation of the fit. You can observe all the multipliers are points on the graphs below

%We assume that investors are risk-neutral, so that only the growth in cashflows matters to them, not their volatility. We focus our attention to contracts of duration 10 years or less. Multipliers are given by

\begin{figure}
\centering
\includegraphics[width=0.7\textwidth]{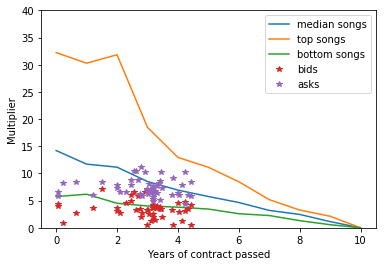}
\caption{\label{fig:market} Multipliers are decreasing in the duration of the contact}
\end{figure}

We observe in Figure \ref{fig:market} that bids are close to the bottom 10th percentile of songs. The asks are close to those justified by a median song. These low values seem to imply buyers assume that the cashflows will decay fast, while sellers assume that they will remain stable. This gap between the bids and the asks may be justified by the inherently asymmetric information between the buyer and the seller.

\begin{figure}
\centering

\includegraphics[width=0.7\textwidth]{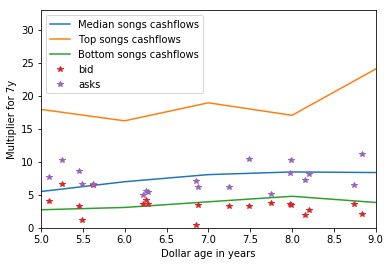}
\caption{\label{fig:age} Multipliers are somewhat increasing in the dollar age of songs}
\end{figure}

We observe in Figure \ref{fig:age} that the dollar age has a very modest effect on the multipliers. Consistently with the behavior in Figure \ref{fig:market}, bids are near the bottom decile and asks are near the median songs.

\pagebreak

\section{Conclusion}

We show that a discounted cashflow model estimated on historical revenue data on Royalty Exchange fit the market data quite well. We find that the ask multipliers are near the value justified by median song cashflows, while the bid multipliers are close to the bottom decile. This market outcome resembles the work of \cite{akerlof} which examines how the quality of goods traded in a market can degrade in the presence of information asymmetry between buyers and sellers. In particular, the seller of music rights may have information that is not available to the buyer, such as the costs associated with the success of songs in a catalog. The mixture of assets that have had organic growth and those that have had significant marketing budgets may make buyers reluctant to buy music rights at their median value.

We see three main directions in which this research could be developed further:
\begin{enumerate}
\item Accounting for risk-averse investors
\item Adjusting multiplier based on other factors than dollar age
\item Adding more revenue data sources for more accurate calculations
\end{enumerate}
The first concern may be addressed by taking a Markowitz optimal investment approach, where both means and variance impact the pricing formula.  %ratio - a common factor used in the finance world to compare investments of different risks. By doing so, we will build a bridge between traditional investment strategies and music assets. Since music assets are very not widely available for portfolios of casual investors, the Sharpe ratio may reveal a vast set of good assets that will attract new buyers.

The two last concerns may be addressed by studying a larger set of songs and their associated streaming cashflows. %If more data is available, we can create more robust models. Preferable data would be revenue streams for musical assets, but due to the 
Since we expect a strong correlation between revenue and streaming counts, historical song streaming data may yield less biased and tighter ranges of reasonable multipliers. 
%As long as we can get the age of the song, any time series of more than 7 quarters of revenues/streams can be incorporated into our model. 

\printbibliography

%\bibliographystyle{alpha}
%\bibliography{sample}

\end{document}